# Ferroelectricity and ferromagnetism in VOI$_2$ monolayer: the role of Dzyaloshinskii-Moriya interaction

Ning Ding,[1] Jun Chen,[1] Shuai Dong[1,*] and Alessandro Stroppa[2,†]

[1]School of Physics, Southeast University, Nanjing 211189, China

[2]CNR-SPIN, c/o Department of Physical and Chemical Science, University of L'Aquila,

Via Vetoio - 67100 - Coppito (AQ), Italy

**Abstract**

Multiferroics with intrinsic ferromagnetism and ferroelectricity are highly desired but rather rare, while most ferroelectric magnets are antiferromagnetic. A recent theoretical work [Phys. Rev. B **99**, 195434 (2019)] predicted that oxyhalides VO$X_2$ (*X*: halogen) monolayers are two-dimensional multiferroics by violating the empirical $d^0$ rule. Most interestingly, the member VOI$_2$ are predicted to exhibit spontaneous ferromagnetism and ferroelectricity. In this work, we extend the previous study on the structure and magnetism of VOI$_2$ monolayer by using density functional theory and Monte Carlo simulation. The presence of the heavy element iodine with a strong spin-orbit coupling leads an effective Dzyaloshinskii-Moriya interaction in the polar structure, which favors a short-period spiral a magnetic structure.. Another interesting result is that the on-site Coulomb interaction can strongly suppress the polar distortion thus leading to a ferromagnetic metallic state. Therefore, the VOI$_2$ monolayer is either a ferroelectric insulator with spiral magnetism or a ferromagnetic metal, instead of a ferromagnetic ferroelectric system. Our study highlights the key physical role of the Dzyaloshinskii-Moriya interaction.

\* Corresponding author. Emails: sdong@seu.edu.cn

† Corresponding author. Email: alessandro.stroppa@aquila.infn.it

**Introduction**

Since the discovery of CrI$_3$ monolayer [1] and Cr$_2$Ge$_2$Te$_6$ few-layers [2] in 2017, two-dimensional (2D) crystals with intrinsic ferromagnetism have attracted a lot of attention boosting both experimental and theoretical research. New 2D ferromagnets have been experimentally confirmed, including VSe$_2$ monolayer [3] and Fe$_3$GeTe$_2$ monolayer [4], and even more have been predicted [5-11]. At the same time, 2D ferroelectric materials have also becoming booming since the discovery of SnTe monolayer [12] and CuInP$_2$S$_6$ few-layer [13] in 2016.

An interesting topic is the crossover of 2D magnetic materials and polar materials, *i.e.*, 2D multiferroics, which represents a new born field of research. In the past decades, the multiferroics in three-dimensional crystals have been extensively studied, but has not widely extended to the 2D families [14-18]. Only till very recently, some 2D materials have been predicted to be multiferroic [19-23]. Not only the type-I multiferroics but also the type-II multiferroics has been designed such as Hf$_2$VC$_2$F$_2$ monolayer with Y-type noncollinear spin texture [19]. Very recently, Tan *et al.* predicted that oxyhalides VO$X_2$ (*X*: halogen) monolayers are two-dimensional multiferroics by violating the empirical "$d^0$ rule" which is a main driving force for proper ferroelectricity as in BaTiO$_3$ [24]. The most interesting member is VOI$_2$, which is predicted to have ferromagnetic (FM) and ferroelectric (FE) orders [21], a very rare but highly desired property.

In the present work, we extend the previous study about VOI$_2$ monolayer [21] by considering both the spin-orbit coupling (SOC) and Hubbard-*U* correction, which were not taken into account in Ref. [21]. In this system, due to the presence of heavy element I, the SOC effect should influence the structural and magnetic properties. For 3*d* orbitals of V, the Hubbard correlation should also be considered. Although in the spin-polarized density functional theory (DFT) calculation, the role of Hund coupling has been partially considered, in practice an additional *U* remains is needed in many cases. Indeed, our calculations find that the combined effect of both SOC and Hubbard-*U* correction is crucial for assessing the ferromagnetism and ferroelectricity in VOI$_2$ monolayer.

**Computational methods**

Our first-principles calculations were performed on the basis of spin-polarized DFT implemented in the Vienna *ab initio* simulation package (VASP) code [27,28]. For the exchange-correlation functional, the PBE parametrization of the generalized gradient approximation (GGA) was used [29] and the Hubbard *U* was applied using on the Dudarev parametrization [30]. In addition, the Heyd-Scuseria-Ernzerhof (HSE06) functional [31] is also adopted to compare with the GGA+$U$ result. The SOC was considered in all calculations, including the structural relaxation. The energy cutoff is fixed at 600 eV, and the V's $2p3d$ electrons were treated as valence states. The $k$-point grid of 11×11×1 is employed to sample the Brillouin zone for the minimal unit cell and accordingly reduced for supercells. A criterion of 0.005 eV/Å is used for the Hellman-Feynman forces during the structural relaxation and the convergence criterion for the energy is $10^{-6}$ eV. A vacuum layer of 17 Å is added to avoid the interaction between monolayer and its periodic images. The polarization was calculated by the standard Berry phase approach [32].

To complement the DFT calculations, the Markov-chain Monte Carlo (MC) method with Metropolis algorithm was employed to simulate the magnetic ordering. The MC simulation was done on a 45×45 lattice with periodic boundary conditions and larger lattices were also tested to confirm the physical results. The initial $1\times10^5$ MC steps were discarded for thermal equilibrium and the following $1\times10^5$ MC steps were retained for statistical averaging of the simulation. The quenching process was used for the temperature scanning. To characterize the magnetic phase transitions, the specific heat $C$ was calculated to determine the critical temperature.

**Results and discussion**

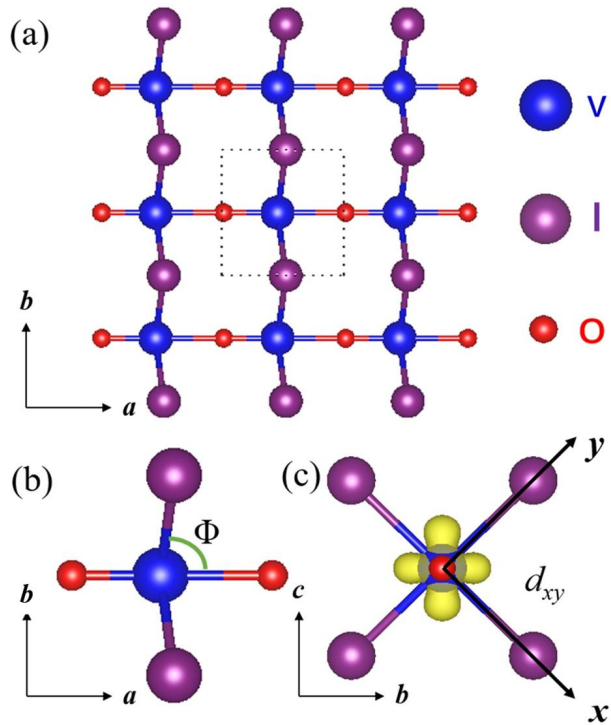

**FIG. 1.** (a) The top view (upper) of the VOI$_2$ monolayer. The dashed rectangle indicates the primitive cell. (b) The side views of a unit cell, where the O-V-I bond angle Φ can characterize the polar distortion. (c) The charge density profile indicates the occupied $d_{xy}$

orbital. Noting that the *x* and *y* directions are defined along the two equivalent V-I bonds in the paraelectric *Pmmm* phase, and the *z* direction is along the V-O bond (*i.e.,* the *a*-axis).

The structure of orthorhombic VOI$_2$ monolayer with the space group No. 25 *Pmm*2 was optimized, as depicted in Fig. 1(a). In such a polar structure, the V cation leaves the center of the octahedral cage defined by iodine and oxygen ions, and the magnitude of such distortion can be qualitatively described by the O-V-I bond angle Φ, as defined in Fig. 1(b). Our optimized lattice constants are *a*=3.797 Å and *b*=3.950 Å for pure GGA calculation with the FM order, which agree well with the previous study [21].

The V$^{4+}$ ion has the 3$d^1$ electronic configuration and the single *d* electron occupies the $d_{xy}$ orbital to avoid the overlap O$^{2-}$ and I$^-$, as illustrated in Fig. 1(c). Thus, along the O-V-O bond, *i.e.,* the *a*-axis (*z*-direction), there is a tendency to form the coordinate bonding, as occuring in those $d^0$ ferroelectrics (like BaTiO$_3$). Thus, the ferroelectricity here is a proper one, instead of improper one. In other words, although V$^{4+}$ is non-$d^0$, it behaves like $d^0$ along the *z*-direction since the orbitals $d_{yz}$ and $d_{xz}$ are empty. Such special anisotropic orbital ordering is responsible for the violation of the $d^0$ rule, *i.e.,* the appearance of proper ferrolectricity, which is coined as "anisotropic $d^1$ rule" here. The estimated 2D in-plane ferroelectric polarization is 225 pC/m (corresponding to 30 μC/cm$^2$ in the three-dimensional case if the thickness of monolayer 7.471 Å is used) along the *a*-axis for the pure GGA calculation with FM order, in agreement with the previous calculation [21].

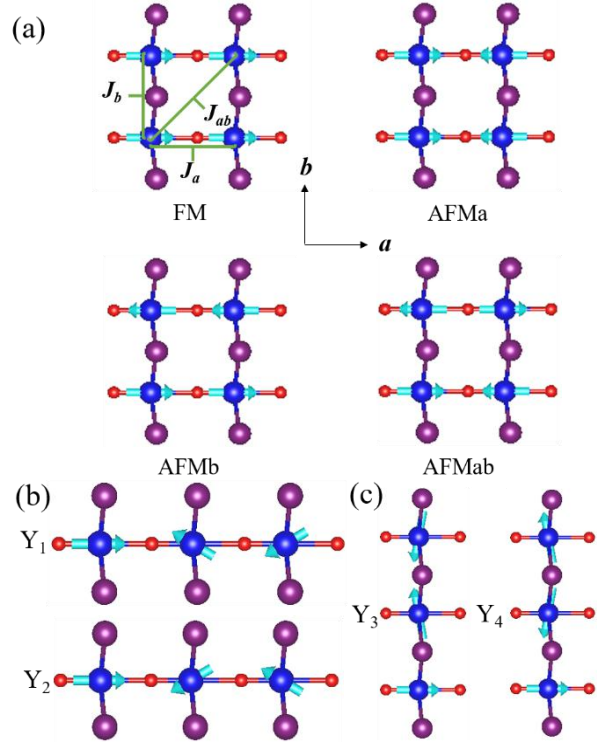

**FIG. 2.** (a) The four collinear magnetic orders, which are calculated to extract the values of $J_a$, $J_b$, and $J_{ab}$. (b-c) The two 120° noncollinear magnetic orders, which are calculated to extract the Dzyaloshinskii-Moriya interactions along $a$-/$b$-direction respectively.

The narrow $3d^1$ band leads to a local magnetic moment up to 1 $\mu_B$/V. The magnetic anisotropy is crucial to stabilize a long-range magnetic order in two-dimensional limit due to the Mermin-Wagner restriction [33]. Using the pure GGA and FM order, the calculated magnetic anisotropy energies were shown in Table I. It is clear that the easy axis of $VOI_2$ monolayer is along the $a$-axis. Then, the exchange interactions between nearest-neighbor (NN) and next-nearest-neighbor (NNN) V pairs were calculated. The NN exchange along the $a$- and $b$-axes are $J_a$ and $J_b$ respectively and the NNN exchange coupling parameter is $J_{ab}$, as depicted in Fig. 2(a). By comparing the energies of four collinear magnetic orders [see Fig. 2 (a)], the values of $J_a$, $J_b$, and $J_{ab}$ can be derived, as shown in Table I. All these three exchange paths prefer the FM coupling and the magnitude of $J_a$ is dominant while $J_b$ is rather weak.

Although all above results agree with those in Ref. [21], supporting the claim of FM ferroelectricity. However, it should be noted that the polar structure breaks the spatial inversion symmetry, which may give rise to the antisymmetric Dzyaloshinskii-Moriya interaction [34,35]. The Hamiltonian form of Dzyaloshinskii-Moriya interaction can be expressed as: $\mathbf{D}_{ij}\cdot(\mathbf{S}_i\times\mathbf{S}_j)$ [34,35], where **D** is a bonding-dependent vector and **S** is spin vector. And due to the heavy element iodine, a strong SOC is expected, *i.e.,* the

Dzyaloshinskii-Moriya interaction should be non-negligible. The interplay of these two effects, i.e., SOC and Dzyaloshinskii-Moriya interaction were not considered in previous study. This is the main motivation of the present work.

Considering the symmetries of such a polar distorted structure, the Dzyaloshinskii-Moriya vector $\mathbf{D}_b$ for the bending V-I$_2$-V bond long the *b*-axis should be along the *c*-axis, while the vector $\mathbf{D}_a$ for the straight V-O-V bond (without the inversion symmetry) should be along the *a*-axis. The magnitude of $\mathbf{D}_b$ should be in proportional to the V-O-V bond bending, *i.e.*, the polarization, and the magnitude of $\mathbf{D}_a$ should be much smaller considering the shape of $d_{xy}$ orbital ordering and the type of symmetry breaking.

By mapping the DFT energy to aforementioned Hamiltonian, these Dzyaloshinskii-Moriya interactions can be calculated based on 3×1×1 and 1×3×1 supercells with two spiral-spin configurations with opposite chirality. The noncollinear spin angles between two neighboring V$^{4+}$ sites were set as 120°, as shown in Fig. 2(b-c), which own different energy contribution from Dzyaloshinskii-Moriya interaction but identical energy contribution from others. The calculated values are summarized in Table I. It is clear that only the *c*-component of $\mathbf{D}_b$ is non-zero, while the *a*-component of $\mathbf{D}_a$ is too small (<0.01 meV). We also checked the source of such $\mathbf{D}_b$ by simply replacing I to Cl in this polar structure, then the magnitude of $\mathbf{D}_b$ becomes (0,0,0.008) meV. Therefore here the heavy element iodine indeed contributes mainly to the SOC-induced Dzyaloshinskii-Moriya interaction.

**Table I.** The DFT calculated magnetic coefficients (meV) for the spin model. The spin is normalized to unit one. Spin-polarized GGA with SOC is adopted. For the Dzyaloshinskii-Moriya vectors, those components below 0.01 meV are considered as zero.

| $K_b$ | $K_c$ | $J_a$ | $J_b$ | $J_{ab}$ | $D_a$ | $D_b$ |
|---|---|---|---|---|---|---|
| 0.11 | 0.54 | -2.15 | -0.69 | -0.72 | (0, 0, 0) | (0, 0, 0.89) |

Then, a classical spin model Hamiltonian can be constructed as:

$$H = J_a \sum_{<i,j>_a} \mathbf{S}_i \cdot \mathbf{S}_j + J_b \sum_{<m,n>_b} \mathbf{S}_m \cdot \mathbf{S}_n + J_{ab} \sum_{\langle\langle k,l \rangle\rangle} \mathbf{S}_k \cdot \mathbf{S}_l + \mathbf{D}_a \cdot \sum_{<i,j>_a} \mathbf{S}_i \times \mathbf{S}_j \\ + \mathbf{D}_b \cdot \sum_{<m,n>_b} \mathbf{S}_m \times \mathbf{S}_n + \sum_i [K_c(S_i^z)^2 + K_b(S_i^y)^2] \quad (1)$$

where $S_i$ is the normalized spin (|**S**|=1) at site *i*; < >*a/b* denotes to the NN along the *a/b*-axis; <<>> represents the NNN along the diagonal direction; $K_{b/c}$ stands for the single-ion anisotropy coefficient.

Based on above DFT coefficients and Hamiltonian (Eq. 1), a Monte Carlo simulation was used to simulate the magnetic ordering of VOI$_2$ monolayer. According to the heat capacity [Fig. 3(a)], there is a peak at $T_C \sim 21$ K indicating a magnetic phase transition. The MC snapshot below $T_C$ confirms a spiral order, as shown in Fig. 3 (b). The spins rotate in the *ab* plane and the propagation vector of spiral is along the *b*-axis. The period or magnetic spiral is ~15 u.c. (about 6 nm) according to the MC simulation [see Fig. 3(b)].

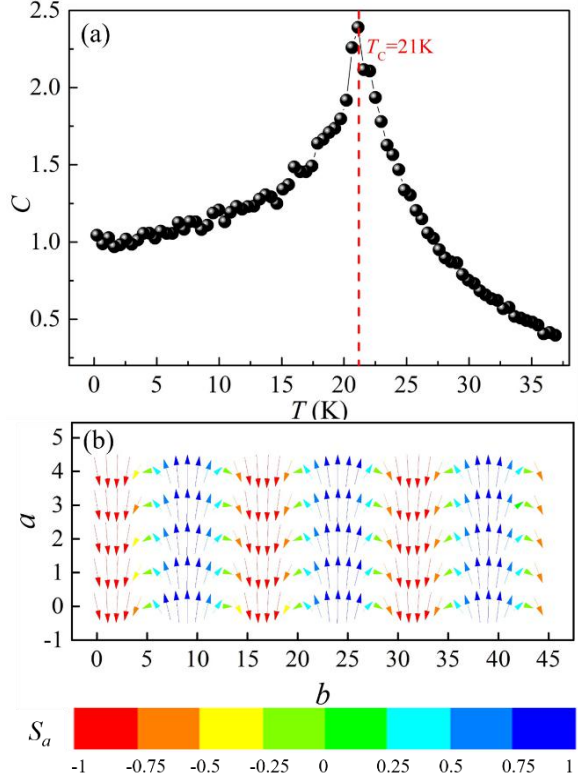

**FIG. 3**. (a) The MC simulated heat capacity $C$ as a function of temperature for the VOI$_2$ monolayer. (b) The MC snapshot of spiral spin order at low temperature. The color bar denotes the *a*-component of normalized spin.

Above results based on pure spin-polarized GGA+SOC have confirmed the ferroelectricity but ruled out the ferromagnetism. Then it is necessary to double-check the Hubbard $U$ correction, which may affect the electron structures seriously especially for partially occupied 3*d* orbitals. In the following, the spin-polarized GGA+$U$+SOC calculations were performed for the structural relaxation, as shown in Fig. 4(a). With increasing $U_{\text{eff}}$ (=$U$-$J$ as defined in the Dudarev approach [29]), the lattice shrinks in the *a*-axis but elongates in the *b*-axis. Such tendency suppresses the polar distortion, as evidenced in the V-O-I bond angle Φ [see Fig. 4 (b)]. When $U_{\text{eff}}$ > 0.4 eV, the polar distortion completely disappears (Φ=90º) and the space group becomes *Pmmm* (*i.e.,* the paraelectric state). The spin-polarized GGA+$U$ calculations without SOC show similar tendency but the critical $U_{\text{eff}}$ are larger, as compared in Fig. 4.

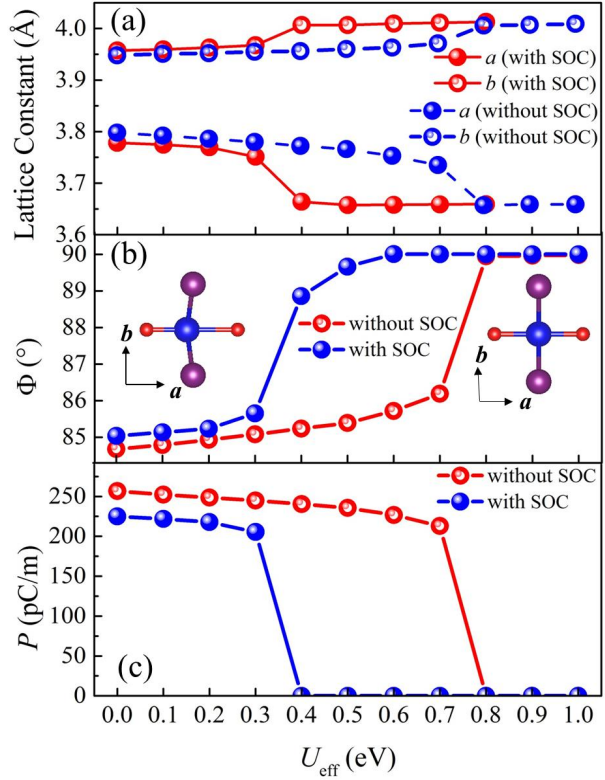

**FIG. 4.** Structural and polarization as a function of $U_{eff}$, calculated with SOC and without SOC. (a) Lattice constants $a$ and $b$. (b) The O-V-I bond angle Φ. Insert: the top view of a unit cell. (c) Ferroelectric polarization.

The electronic structure is also sensitive to $U_{eff}$. The density of states (DOS) of VOI$_2$ monolayer calculated by GGA+SOC and GGA+$U$+SOC ($U_{eff}$=1 eV) were shown in Fig. 5(a) for comparison. Surprisingly, the VOI$_2$ monolayer turns to be metallic when $U_{eff}$=1 eV, while originally it is a semiconductor. This tendency is not usual, in opposite to the empirical rule of DFT+$U$ calculation which prefers the Mottness (i.e. to open a band gap). The band gaps for both spin channels are summarized in Fig. 5(b). The effective band gap is determined by the spin-up channel, which disappears at $U_{eff}$~0.4 eV, in consistent with above structural transition.

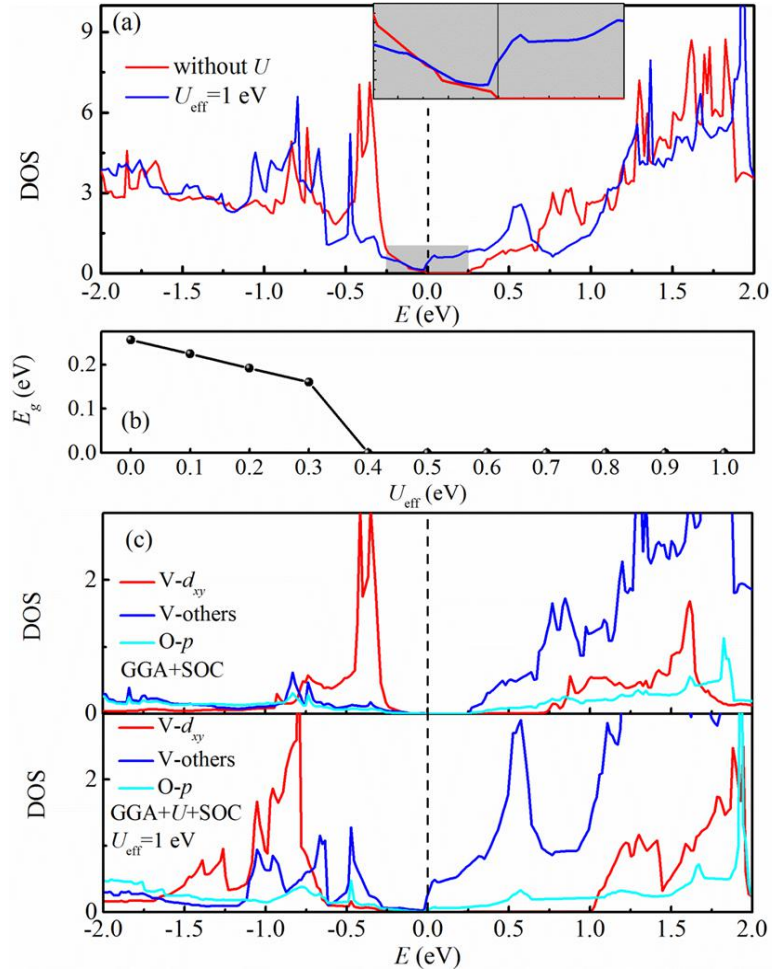

**FIG. 5.** (a) Comparison of DOS's of VOI$_2$ monolayer with/without $U_{\text{eff}}$. Insert: magnified view near the Fermi level. (b) The band gaps of spin-up and spin-down channels as a function of $U_{\text{eff}}$. (c-d) Projected DOS (PDOS) of V's $d$ orbitals and O's $p$ orbitals with/without $U_{\text{eff}}$. Two characteristics are clear for the $U_{\text{eff}}=1$ eV case. First, the Hubbard splitting is enlarged. Second, the occupied $d^1$ bands are broader.

Such strongly $U$-driven insulator-metal transition and polar-nonpolar transition can be understood as following. The polar distortion is induced by the anisotropic $d^1$ condition as discussed before, *i.e.,* the formation of coordination bond between V$^{4+}$ and O$^{2-}$, which relies on the orbital hybridization between V$^{4+}$'s empty $d_{xz}/d_{yz}$ orbitals and O$^{2-}$'s occupied $p_z$ orbitals. However, the inter-orbital Hubbard repulsion will push the $d_{yz}/d_{xz}$ orbitals to upper energy region [see Fig. 5(c)], which suppresses the $d_{xz}/d_{yz}$-$p_z$ orbital hybridization considering the larger energy gap. Therefore, the polar distortion is suppressed by increasing $U_{\text{eff}}$ as found in Fig. 4(b). As a result of such polar distortion suppression, the crystal field splitting between the $d_{xy}$ and $d_{yz}/d_{xz}$ orbitals is reduced. Then the bandwidth of valence band becomes broader, since more and more $d_{yz}/d_{xz}$ components are mixed in. Finally, this VOI$_2$ monolayer becomes a metal.

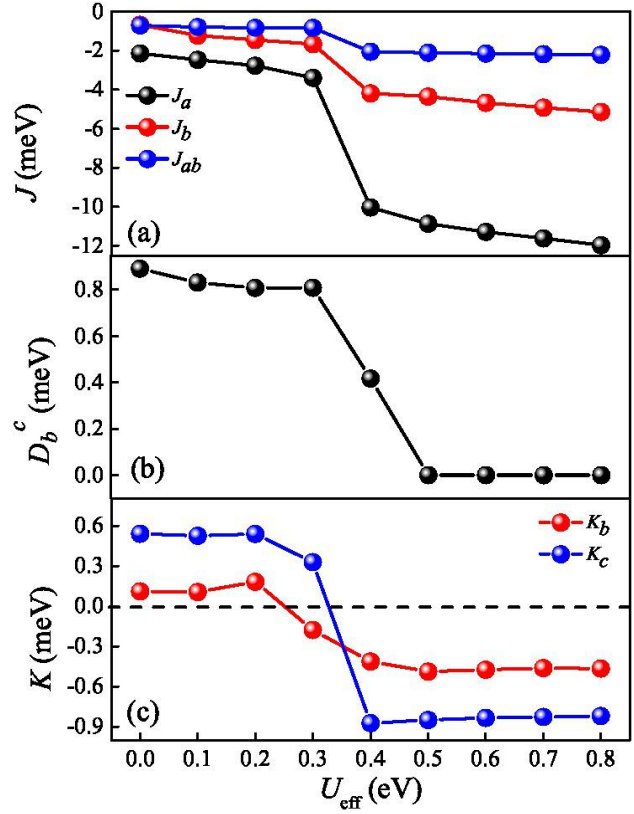

**FIG. 6.** Magnetic coefficients as a function of $U_{eff}$. (a) The exchange interactions. (b) The $c$-component of $\mathbf{D}_b$. (c) The magnetocrystalline anisotropic coefficients.

Since both the crystal structure and electronic structure of $VOI_2$ monolayer change with $U_{eff}$, the magnetic properties should also be sensible to $U_{eff}$. The exchanges and Dzyaloshinskii-Moriya interactions are re-calculated as a function of $U_{eff}$. As shown in Fig. 6(a), all exchanges $J_a$, $J_b$, and $J_{ab}$ are enhanced with increasing $U_{eff}$, especially at the transition point from ferroelectric state to paraelectric metallic state. The physical reason can be understood as following. First, the shrinking of lattice constant along the $a$-axis strengthens the exchange $J_a$. Second, the expanding of lattice constant along the $b$-axis makes the V-$I_2$-V more straight while the V-I bond length does not change too much, which strengthens the superexchange $J_b$. Also the metallicity will enhance the itinerant of electrons, which will strengthen the long-range exchange $J_{ab}$. While for the Dzyaloshinskii-Moriya interaction, as shown in Fig. 6(b), it changes following the behavior of polar distortion [Fig. 4(b)] since it is directly determined by the symmetry. In particular, for the nonpolar structure, $\mathbf{D}_b$ becomes zero. With the increasing exchange interactions and decreasing Dzyaloshinskii-Moriya interaction, the spin-spiral period of $VOI_2$ monolayer will becomes longer and longer, and finally the system becomes a ferromagnetic metal.

Finally, the HSE06+SOC approach has been used to verify above GGA+$U$+SOC

calculation. The HSE06+SOC optimization leads to even stronger polar distortion (e.g. Φ=83.54°) than the GGA+SOC result (Φ=85.03°). Correspondingly, the polarization obtained in the HSE06+SOC calculation is 258 pC/m, which is larger than that of GGA+SOC (225 pC/m). Then the stronger Dzyaloshinskii-Moriya interaction and spiral spin texture is expectable.

**Conclusion**

The structural, electronic properties, electric polarization, as well as magnetic property of $VOI_2$ monolayer have been studied systematically via spin-polarized GGA+SOC and GGA+$U$+SOC methods. Our results have confirmed but, at the same time, go beyond the previous work. In particular, our work revealed the key role of antisymmetric Dzyaloshinskii-Moriya interactions which is significant in $VOI_2$ monolayer. Our conclusion is that $VOI_2$ monolayer is either a ferroelectric magnet with spiral spin configuration (in the low $U$ limit or using the HSE method), or a ferromagnetic metal without ferroelectricity (in the large $U$ side), instead of the expected ferroelectric ferromagnet. Our work will stimulate future experimental verifications and suggest more comprehensive considerations when searching for new 2D multiferroic materials.

Note added. Recently, we became aware of a recent theoretical work on $VOI_2$ monolayer [36], which reported a similar noncollinear spin order as ground state.

Work was supported by National Natural Science Foundation of China (Grant Nos. 11834002 and 11674055). Most calculations were done on Tianhe-2 at National Supercomputer Centre in Guangzhou and the Big Data Computing Center of Southeast University.